# A Hybrid Multi-GPU Implementation of Simplex Algorithm with CPU Collaboration

Basilis Mamalis [1], Marios Perlitis [2]

[1] *University of West Attica, Agiou Spyridonos, 12243, Egaleo, Athens, GREECE*
[2] *Democritus University of Thrace, University Campus, 69100, Komotini, GREECE*

[1]vmamalis@uniwa.gr
[2]mperlitis@gmail.com

*Abstract*—The simplex algorithm has been successfully used for many years in solving linear programming (LP) problems. Due to the intensive computations required (especially for the solution of large LP problems), parallel approaches have also extensively been studied. The computational power provided by the modern GPUs as well as the rapid development of multicore CPU systems have led OpenMP and CUDA programming models to the top preferences during the last years. However, the desired efficient collaboration between CPU and GPU through the combined use of the above programming models is still considered a hard research problem. In the above context, we demonstrate here an excessively efficient implementation of standard simplex, targeting to the best possible exploitation of the concurrent use of all the computing resources, on a multicore platform with multiple CUDA-enabled GPUs. More concretely, we present a novel hybrid collaboration scheme which is based on the concurrent execution of suitably spread CPU-assigned (via multithreading) and GPU-offloaded computations. The experimental results extracted through the cooperative use of OpenMP and CUDA over a notably powerful modern hybrid platform (consisting of 32 cores and two high-spec GPUs – Titan Rtx and Rtx 2080Ti) highlight that the performance of the presented here hybrid GPU/CPU collaboration scheme is clearly superior to the GPU-only implementation under almost all conditions. The corresponding measurements validate the value of using all resources concurrently, even in the case of a multi-GPU configuration platform. Furthermore, the given implementations are completely comparable (and slightly superior in most cases) to other related attempts in the bibliography, and clearly superior to the native CPU-implementation with 32 cores.

## I. INTRODUCTION

Linear programming is the most known and well-studied optimization problem. The simplex algorithm found in many textbooks [1] has been successfully used for many years to solve linear programming problems. Parallel approaches have also been extensively studied because they require intensive computations and require much faster implementations that make effective use of modern computer architectures.

Most studies (related to the sequential simplex) focus on the revised method because of the rarity inherent in most linear programming applications. The revised method is also useful for problems with high aspect ratios. That is, for problems with more columns than rows. However, there have been few parallel / distributed implementations of highly extensible and revised methods [2]. On the other hand, the standard method for dense linear problems is more efficient and it can be easily converted to a distributed / parallel version with satisfactory performance and good scalability [2-5]. A detailed overview can be found in Section II. Recently, some other very promising alternatives have also been communicated, based on the block angular decomposition (or native structure) of the problem in its initial transform [6-7], which have pointed out excessive performance for LP problems of large and very large size in a multi-core distributed memory environment.

Moreover, considering parallelization, the relevant tools, models and libraries for shared and distributed memory platforms, have evolved separately until recently, highlighting MPI library [8] as the leading programming tool for the distributed memory approach, and OpenMP API [9] as the most suitable for thread-based programming with shared memory. Recently, the hybrid model (i.e. through the combined use of MPI and OpenMP) has been in the spotlight for more than one reasons. Firstly, it's relatively easy to choose a language or library instance for the hybrid model itself. Secondly, the popularity of the model as well as the popularity of scalable parallel computers is outstanding nowadays. Almost all of today's fastest machines (including supercomputers) consist of a large number of multi-core nodes connected by a high-speed network. The most recommended option is to use OpenMP threads (using multithreaded processes per node) to utilize multiple cores per node while using MPI for communication between nodes. The MPI shared memory support mechanism is also a good alternative for efficient parallelization within each multi-core node [10,11,18].

Furthermore, the computational power provided by the massive parallel processing of modern graphics processing units (GPUs) is now increasingly attracting attention for several types of GPU-accelerated solutions. General purpose graphic processing units (GPGPUs) have evolved enormously the last decade, being considered nowadays as an essential component in almost all supercomputers (see website https://www.top500.org). Simplex parallelization hasn't seen as many decent trials as expected, and although parallel GPU based simplex algorithm implementations have not yet delivered notably better performance than efficient sequential simplex solvers in most cases, significant progress has been made, at least for high-density LP tasks [12]. There are also different approaches in the field of parallel scientific computing using extended GPU/CPU cooperation [13-17]. However, efficient GPU/CPU concurrent operation through the combined use of appropriate programming models (e.g. OpenMP and CUDA) remains a major research challenge. Also, to the best of our knowledge, there is no adequate approach in the literature to parallelize the simplex approach using GPU/CPU interaction using multiple GPUs.

This article gives special focus on parallelizing the standard full tableau simplex method. In particular, it presents a novel, highly efficient, parallel implementation of the standard simplex method that simultaneously utilizes all the features of an underlying multicore hardware platform, along with multiple GPUs with CUDA support. The computing resources of the proposed platform are configured in a novel hybrid cooperative approach based on the combined execution of multi-threaded computations and offloaded GPU operations, that are suitably balanced. Also, two high-performance GPUs (Titan Rtx and RTX 2080Ti) are utilized to test the high performance of the proposed hybrid approach. All experiments were performed on an appropriate subset of the well-known and widely used NETLIB LP test questions and on randomly generated large-scale and extra-large LP problems.

The results obtained through the above experiments show that the performance of the proposed GPU/CPU cooperative approach outperforms the existing high-performance GPU-accelerated solution under almost all conditions, confirming the value of using both resources at the same time. The GPU-only implementation is comparable to other related works in the literature, and performs significantly better than the CPU-only implementation. More specifically, the GPU-only implementation produces speedup values of up to 2.34, 2.76, and 4.13, respectively (versus the 32-core CPU-only implementation – when using the Rtx 2080Ti, Titan Rtx, and both of them in cooperation). The hybrid GPU/GPU implementation also results in additional speedup values of up to 1.29, 1.24 and 1.13, respectively (compared to the GPU-only implementations – when using the Rtx 2080Ti, Titan Rtx, and both of them in cooperation). The above results are very satisfactory and comparable to the results presented in other studies in the recent literature [11,14,16-17,20]. They are also the first to achieve such high speedup values with a multi-GPU configuration (i.e. under hardware conditions that make the CPU component less competitive).

The remaining of the paper is organized in the following sections. The most recent related work is stated in Section II. A brief overview of the standard simplex method is given in Section III. In Section IV our hybrid parallelization approach (based on GPU/CPU collaboration) is described in details. Our extensive experimental evaluation is presented in Section V, whereas in Section VI the conclusions of the paper are drawn.

## II. Related Work

Most of the research attempts made in the earlier years with respect to simplex parallelization mainly focused on shared memory environments as well as on tightly coupled platforms and clusters. In [21-22] the authors present two of the very first efficient parallelization schemes of the revised simplex method over powerful parallel environments of shared memory and hypercube-based interconnection patterns. On the other hand, in [23] Thomadakis & Liu present a quite effective parallel implementation of the standard simplex method utilizing the MP-1 and MP-2 MasPar. Furthermore, in [24] Eckstein et al. – working on the parallel connection machine (CM-2) – show that the iteration processing time of the parallel implementation of the revised simplex method is higher than the one achieved for the implementation of the full tableau standard method. Also, in [25] Stunkel presents a parallelization scheme for both the above methods (standard and revised), which led to similar speedup values when implemented over the parallel (based on hypercube interconnection network) Intel iPSC supercomputer. Additionally, in [26-27] two other worth-telling attempts are presented, which focus on parallelizing the primal-dual simplex method and the sparse simplex method, and they both lead to competitive results for large linear problems.

Considering the most recent attempts, the one presented by Huangfu and Hall [28-29] is probably regarded as the most significant one. In [28-29], the authors present the design and implementation of a notably scalable and efficient parallel scheme of the revised simplex method (in its dual form), following the technique of suboptimization, and they have obtained very satisfactory response times, which are highly competitive with the ones given by the best commercial solvers. The reader may find a more detailed presentation of the most important research works in simplex parallelization, in [12].

As it has already been stated above, the standard simplex method can be easily decomposed to dominating independent tasks, and consequently it can lead to effective parallelization schemes over either shared or distributed memory environments [5]. In addition, the dense linear problems (which favor the use of the standard method) tend to occur quite frequently in several important applications in linear programming [24] (although they are considered rather uncommon in general). Furthermore, the parallel approaches of the standard simplex method met in the bibliography (over distributed memory environments), usually vary according to the distribution pattern of the tableau over the multiple processors [1,30]. The use of either a global row distribution scheme or a column distribution scheme may be chosen, based on specific parameters (such as the actual number of rows and columns, the size of the problem, the details of the hardware platform etc.). The column distribution approach has been followed in many related research attempts in the past, such as in [5] by Yarmish et al. and earlier in [4] by Qin et al. As an alternative, Badr et al. present in [3] a quite efficient implementation following the row distribution scheme over a loosely connected distributed-memory platform. The two above different data distribution approaches have also been thoroughly studied by Mamalis et al. in [18,30] and the relevant implementations have been extensively tested, leading to particularly high speedup and efficiency values. Eventually, the column distribution approach is shown to be more efficient in the most experimental cases (especially when large and very large-scale LP problems are considered in the evaluation process).

Moreover, a lot of valuable attempts for parallelizing simplex have been made with the use of GPUs. In [31], Spampinato et al. present an efficient parallelization scheme for the revised simplex method, which is based on the well-known libraries CUBLAS and LAPACK, and leads to a maximum speedup of 2.5 with the use of a GTX 280 GPU. The corresponding results have been obtained versus a sequential implementation on a computer with Intel Core2 Quad 2.83 GHz, and randomly generated linear programming problems of size 2000x2000. In [32], the authors introduce another valuable GPU implementation of the revised simplex algorithm, in which a maximum speed up of 18 (with single precision computations) is achieved, with the use of a GeForce 9600 GPU card (versus a GLPK solver running on a workstation with a 3GHz Intel Core 2 Duo CPU).

In [19], Lalami et al. present and evaluate a parallel implementation of the standard simplex algorithm, which is mainly based on GPU acceleration via CUDA, targeting to dense LP problems. The corresponding experiments were made with the use of a 3GHz Intel Xeon CPU and a GTX 260 GPU, and they have led to significant speedup values up to 12.5 (in double precision), over randomly generated test problems with sizes up to 8000x8000. The proposed approach has also been generalized over a multi-GPU platform, thus leading to a maximum speedup of 24.5 when evaluated over two C2050 Tesla boards. In [33], Meyer et al. present two versions of a direct implementation of the standard simplex method (using one and multiple GPUs respectively), and they further compare their performance to the sequential Clp solver, over a S1070 Tesla board with T10 GPUs.

The proposed implementations led to quite better response times than the Clp solver on large LP problems. Finally, in [35], the authors propose two notably efficient implementations of the revised simplex method and the exterior point method, over a high-end GPU platform utilizing Matlab. According to their experimental results, the implementation of the exterior point method led to quite high speedup values, whereas the speedup achieved for the revised simplex method was very poor. The reader may also find some other quite valuable approaches in [36-38], which have led to acceptable speed-up values with the use of C1060, S1070 and GTX 670 GPUs and supposing double precision computations.

In [39] the authors present a novel generic framework that transparently orchestrates collaborative execution of a single data-parallel kernel across multiple asymmetric CPUs and GPUs. Also, in [40] the authors present a quite promising collaboration approach in which the intensive matrix manipulations of traditional revised simplex algorithm (RSA) are offloaded to the GPU, which helps to make full use of the parallel computation capability of GPU to accelerate matrix manipulations of this algorithm. The numerical experiments on randomly generated LPs demonstrated that GPU-based RSA can not only get correct optimal solution, but also reach as fast as hundred times of CPU-based RSA. A similar approach if followed by the authors in [41]. In this work, to decrease the data delay between the GPU and the GPU's memory, the authors designed optimized threads and blocks of NVIDIA's CUDA so the kernel could do intensive arithmetic calculations and conceal the latency. Moreover, to reduce the data transfer overhead, they introduce 2 strategies: (a) they propose the use of very large arrays residing in the global memory of the GPU without being downloaded from the GPU to the CPU, and (b) only two small vectors are downloaded in the iteration in direct memory access (DMA) mode to minimize data traffic.

Also, a quite interesting study is presented in [42] with regard to the case of Mixed Integer Programming (MIP) problems. The authors note there that the best potential lies in solving problems whose individual matrix sizes (of the linear program relaxation) fit entirely within one accelerator's memory and whose branch-and-bound (or branch-and-cut) trees cannot be fully contained within a small number of computational nodes. In [43], firstly the performance of existing GPU-based Basis Update (BU) techniques is analyzed, and secondly the adapted element-wise technique is extended to develop a new BU technique by using inexpensive vector operations. In addition, in [44] the authors present the design and implementation of a batched LP solver in CUDA, keeping memory coalescent access, low CPU-GPU memory transfer latency and load balancing as the goals. Some interesting new memory management strategies are also presented, analyzed and evaluated in [45] with regard to the parallelization of the RSA, whereas to avoid cycling because of degeneracy conditions, a tabu rule for entering variable selection is used.

Furthermore, considering the wide area of scientific computing, a lot of research efforts have been made adopting extended CPU/GPU collaboration. In [15], the authors present Harmony, an integrated programming framework which allows coding and executing programs over hybrid CPU/GPU computing architectures. The automated distribution of the computational load on both the CPU and GPU components is also included, leading to particularly high performance mainly when used over audio-processing platforms. In addition, in [14,16-17] some more recent related attempts are given in the general area of linear algebra and solving of linear systems. In [17] the authors present a highly effective approach which leads to a speedup of up to 1.25 (comparing to the GPU-only based implementation) for the parallel implementation of the conjugate gradient method. Also, in [14,16] two other analogous implementation schemes (which are totally based on CPU/GPU collaboration) are introduced in the field of linear algebra, which eventually lead to speedup values ranging from 1.15 up to 1.35 for different problem sizes and types. In the works of [14,17] another major advantage is also included; the capability of distributing the workload between CPU and GPU dynamically.

To our knowledge there is no relevant approach in the literature adopting extended CPU/GPU collaboration for parallelizing the simplex method, especially involving the use of multiple GPUs in parallel. A first attempt is presented in our previous works of [11,20], however it considers the case of only one GPU and it has not been tested over high performance GPGPU platforms with recent technological features that enhance that kind of collaboration by appropriately reducing the CPU-to-GPU memory transfer overhead. However, the most important achievement there, is obtained when the proposed scheme is used in combination with MPI in a multi-node environment, which leads to substantial improvements in the speedup achieved for large and very large LP problems. An analogous combination (in a fully hybrid distributed environment) may be also applicable with use of the multi-GPU collaboration scheme presented here, and it could lead to even greater performance gains.

## III. THE STANDARD SIMPLEX METHOD

In linear programming (LP) problems [1,34], the main objective is to minimize / maximize a linear function of real variables over a region defined by linear constraints. With respect to its standard form, the simplex method can be described as given in Table I, where A is an mxn matrix, x is an n-dimensional vector of real variables, c is the PRICE vector, b is the vector of the right-hand side of the constraints, and T denotes transposition. It is also assumed that the set of basis vector (i.e. the A-matrix columns) is linearly independent. Based on this setup, the simplex algorithm consists of the following two main steps: (a) a method of estimating whether a current basic feasible solution can be considered as an optimal solution, and (b) a method for choosing a new (adjacent) basic feasible solution (which has the same or better value for the objective function) for the next iteration. Here, we focus on the standard full tableau representation of the simplex algorithm, which has been proved to be more efficient when applied to full dense linear problems, and also it can be easily turned to an effective parallelized form, suitable for cluster systems as well as for hybrid computing platforms. The possibility of ideal and flexible decomposition of most of the implied computations in independent tasks is the main advantage of the full tableau version when considering efficient parallelization.

TABLE I. SIMPLEX FULL TABLEAU REPRESENTATION

| | $x_1$ | $x_2$ | ... | $x_n$ | $x_{n+1}$ | ... | $x_{n+m}$ | z | |
|---|---|---|---|---|---|---|---|---|---|
| | $-c_1$ | $-c_2$ | ... | $-c_n$ | 0 | ... | 0 | 1 | 0 |
| $x_{n+1}$ | $a_{11}$ | $a_{12}$ | ... | $a_{1n}$ | 1 | ... | 0 | 0 | $b_1$ |
| $x_{n+2}$ | $a_{21}$ | $a_{22}$ | ... | $a_{2n}$ | 0 | ... | 0 | 0 | $b_2$ |
| ... | ... | ... | ... | ... | ... | ... | ... | ... | |
| $x_{n+m}$ | $a_{m1}$ | $a_{m2}$ | ... | $a_{mn}$ | 0 | ... | 1 | 0 | $b_m$ |

Following the above considerations, the main steps of the standard simplex algorithm may be described (without loss of generality) as presented below:

**Initialization Step:** Start having as a basis a feasible basic solution and construct the tableau.

**Step 1:** Choose the entering variable: find the winning column (the column with the larger negative coefficient of the objective function – this is the 'entering' variable).

**Step 2:** Choose the leaving variable: find the winning row (run the minimum ratio test on the items of the winning column and conclude to the row having the minimum ratio – this is the 'leaving' variable).

**Step 3:** Pivoting (the step with the most calculations): form the new simplex tableau (for the next iteration) by applying pivoting in the rows of the previous tableau, using the new pivot row found in step 2.

**Iterate/Finalization Step:** Iterate on steps 1-3 till finding the best solution or the problem is proved to be unbounded.

IV. THE HYBRID MULTI-GPU PARALLELIZATION

In the next paragraphs the algorithmic approach adopted in as the basis of our hybrid parallel schem is presented in details. The proposed approach follows the column-based distribution scheme [4], which is the most efficient and widely adopted in the literature. It implies a rather straightforward way of parallelizing the standard simplex algorithm, in which the columns of the simplex tableau are equivalently spread among all the resources; thus leading to the best performance in the general case. Adopting this approach most of the computing steps (all except step 2) of the sequential algorithm are completely parallelized. In addition, this way of parallelization seems to be the most appropriate choice because in the majority of the practical problems met in the real life, the number of columns is greater than the number of rows. Also, it has been analyzed and evaluated as the most efficient practical approach in several recent research works [5,30].

More concretely, based on the stepwise decomposition given in the previous section, we've designed and implemented our basic hybrid parallelization scheme, assuming a hybrid parallel platform which consists of a heavy *multicore* machine (i.e. with a large number of CPU cores) and a number of modern general purpose GPUs. The *OpenMP* API was used for the shared-memory multiple cores communication within the CPU, and CUDA programming framework was used with respect to GPU computations. Apparently, the proposed scheme may also fit in a straightforward manner to a fully hybrid multi-node computing environment (being the basic parallel working scheme of each fat/heavy single node of the system). Based on the above general considerations, and assuming that in our case we have a single node with

- *n* CPU-cores, and

- *m* CUDA-enabled GPUs,

and (without loss of generality) $n > m+1$, we first suppose (upper-level parallel approach) that a global column-based distribution scheme is followed with regard to the distribution of the simplex tableau among the provided resources. In this context the proposed hybrid CPU+GPU implementation (OpenMP+CUDA) has as follows:

***Setup:*** A process with $t$ ($n > t > m$) threads is scheduled, where $m$ threads are mainly used for GPU handling (i.e. for the offloading and kernel launching procedures), whereas the remaining $t-m$ threads are kept for CPU-assigned computations.

***Initialization:*** A portion $\theta$ (where $\theta \in [0,1]$ and it's appropriately chosen according to the processing capabilities of the present hardware components) of the simplex tableau is decided to be offloaded to the present GPUs, whereas the remaining portion $(1-\theta)$ is kept in CPU's memory for shared computations among the $t-m$ CPU threads.

**Step 0:** The former (portion $\theta$ of the simplex tableau) is further divided/partitioned to $m$ parts $\theta_i$ ($i=1...m$) according to the processing capabilities of the present GPUs (ideally – in the case of m identical GPUs – it's divided to $m$ equal parts), and then each of them is offloaded to the corresponding GPU (i.e. the $\theta_i$ part is offloaded to the $i^{th}$ GPU). Additionally, the right-hand constraints vector (cv) is offloaded to all the present GPUs and it's also kept in the CPU shared memory too.

**Step 1:** The $t-m$ CPU threads and all the $m$ GPUs compute (in parallel) the maximum negative coefficient over their own portion of the first line of the simplex tableau, yielding to $m+1$ local maximum index values. Let's denote $k_i$ ($i=1...m$) the local maximum index values computed by the m GPUs and $k_0$ the one computed by the CPU. The GPUs local maximum index values $k_i$ are then transferred to the CPU's memory, and then they are compared to $k_0$, and the global maximum index ($k$) of the winning column is determined (entering variable).

**Step 2:** If the entering variable belongs to the portion $\theta$ that has been offloaded to the present GPUs (specifically, let's say it belongs to part $\theta_j$ offloaded to the $j^{th}$ GPU), the index $k$ is transferred to the corresponding ($j^{th}$) GPU's memory. Then, the *ratio computation* is applied to all the data of the winning column lying in that ($j^{th}$) GPU. The *minimum ratio* is also computed in parallel in that GPU and the index $r$ of the winning row is specified (namely, the leaving variable). Then, the index $r$ of the winning row and the data of the winning column are sent to the CPU, as well as to the rest of the GPUs.

**Step 3:** If the entering variable belongs to the portion $(1-\theta)$ that has been kept for CPU-assigned computations, the $t-m$ CPU threads apply the *ratio computation* in parallel to all the data of the winning column in the CPU memory. Afterwards, they also compute the *minimum ratio* in parallel, and the index $r$ of the winning row is specified (namely, the leaving variable). Then, the index $r$ of the winning row and the data of the winning column are sent to all the GPUs memories.

**Step 4:** Finally, the $t-m$ CPU threads and all the $m$ GPUs perform in parallel all the computations required for the global rows pivoting, on their own part of the simplex tableau, using the pivot data received during the previous step. This is the step with the most calculations in each iteration, and it's ideally parallelized.

***Iterate/Finalization:*** The above *Steps 1-4* are repeated until the best solution is found or the problem gets unbounded. A suitable synchronization is required in this step between CPU and all the GPUs per iteration.

With regard to the implementation of the above steps, through OpenMP and CUDA, the available functions, constructs, and specialized mechanisms of both the above programming APIs were adequately used as follows:

- Suitably developed OpenMP *parallel for* directives were applied with respect to the effective thread-based parallelization of the loops implied by steps 1, 3 and 4. More specifically, the column distribution scheme was followed to spread the simplex tableau among all the computing components, which has been shown to be the best choice for parallel use.

- Considering the parallel implementation of steps 1 (in combination with step 2) and 3, in order to optimize the parallel execution of the relevant procedures (which both imply a reduction operation), we've used the optimized *min/max reduction operators* (and the related directives) offered by the OpenMP API.

- Moreover, with respect to the parallel implementation of step 4, trying to obtain the proper distribution of calculations to all the participating threads (provided that the calculations costs of the main loop iterations should not be considered as equivalent in the general case) we used *collapse-based nested parallelism* directives in combination with appropriately defined *dynamic scheduling policy*.

- In addition, regarding the parallel implementation of steps 1 and 4 in the participating multiple GPUs, the technology of *cuda streams* has been adequately used in order to efficiently utilize all the computing units in parallel. The required hardware support (*NVLink* or *P2P*) is also assumed to be in use. It should be noted here that for more than two GPUs *P2P* support would be necessary for the most efficient behavior. For two participating GPUs (as in our experiments) an *NVlink* interconnection is sufficient.

- Furthermore, as it was stated earlier, with respect to the proper load distribution required for the effective CPU/GPU collaboration, we apply a column-based distribution pattern of the simplex tableau. However, with respect to the internal processing within the GPU-cores, the corresponding distribution pattern (of the relevant tableau portion) is set to a *block-oriented one*, which is the best choice considering the inside architecture and the processing peculiarities of an NVIDIA GPU [19,30].

It must also be noted that in the proposed hybrid scheme, a proper share of the computations required for the parallel implementation of the min/max operations denoted in the first two steps (1 and 2) of the sequential simplex algorithm, are being executed in the GPUs, using suitable optimized tree-based reduction techniques. As it was shown in our experiments, the performance gained by spreading these reduction operations to both the CPU and the GPU, is clearly faster than the alternative followed in earlier research works (i.e. in [19,20] where the reduction operations are completely performed within the CPU). The peculiarities of the NVIDIA architecture as well as the quite large scale of the test linear problems, lead to rather limited performance efficiency when the GPU is the dominant computing component executing the reduction computations. Considering the recent/modern NVIDIA GPUs however, the efficiency of such computations has been substantially improved, permitting their adequate use in the relevant computing procedures. A more clear presentation of the implementation flow of the proposed hybrid CPU/GPU collaboration approach is given in Fig.1.

## V. EXPERIMENTAL RESULTS

Our hybrid parallelization scheme presented in the previous section has been implemented using OpenMP 5.1 API and CUDA 10.2 Toolkit. It has also been thoroughly tested over a real hybrid hardware platform. More concretely, for our basic set of experiments we've used a powerful Intel Xeon multi-core system of 32 cores in total (2 x Intel Xeon Gold 6142 processors) and 128GB RAM, as well as two modern NVIDIA GPUs (one Rtx 2080Ti/11GB and one Titan Rtx/24GB) which are of the same technology (Turing architecture). These desktop-level GPUs have extremely high single precision (SP) performance and relatively low double precision (DP) performance (the Rtx 2080Ti gives ~13447/420 Gflops SP/DP performance with 4352 cores, whereas the Titan Rtx gives ~16312/509 Gflops SP/DP performance with 4608 cores), however as it can be seen they can lead to quite significant improvements. This emphasizes the capability of using GPUs for scientific computing on desktop environments too.

Note also that we've chosen to use the above referred 32-core Xeon system in order to have sufficiently high number of available cores in a single machine, and conclude to more representative, convincing and reliable results. The relevant computing components (for both the multicore and gpgpu platforms) were mainly accessed through the local infrastructure of the University of West Attica and the Democritus University of Thrace.

### A. Performance of the basic OpenMP scheme

Firstly, in order to evaluate and demonstrate the high effectiveness and scalability of the proposed OpenMP-only based parallel scheme (without GPUs acceleration, i.e. by setting $\theta$ equal to zero: $\theta=0$), we've run on our platform a suitable subset of the well-known and extensively used NETLIB test linear problems (LPs). Specifically, we've used test LPs of variable sizes (large and very large) which are close to the realword practical cases. The experimental results obtained over varying (in powers of two) numbers of cores (from 4 up to 32) are given in Table II. The *speed-up* measure for $p$ cores ($Sp$) is calculated as the execution time needed for one core divided by the execution time needed for $p$ cores, whereas the *efficiency* measure for $p$ cores ($Ep$) is calculated as the *speed-up* obtained in $p$ cores divided by $p$. The *efficiency* measure reflects the percentage of the maximum expected speed-up that has been actually achieved.

As it is shown in Table II the speed-up and efficiency values obtained for the proposed OpenMP-only based scheme are especially high in all cases, even when large number of cores and very large LPs are considered. It can also be easily observed that the efficiency values decrease as the number of cores increase; as it was normally expected because of the increased overhead for threads management. The corresponding decrease is however slow, whereas the efficiency values are still high in all cases, even when 16 and 32 cores are encountered (getting values no less than 83% and 77%, respectively). In addition, especially high values are observed with respect to the efficiency measure (reflecting almost linear speedup) for all the problems with high aspect ratio (e.g. the efficiency values for FIT2P, 80BAU3B and QAP15 test linear problems over 16 and 32 cores are over 91% and 87%, respectively). Note also that the maximum speed-up values (achieved when all the 32 cores of the system are used) range between 24.75 and 30.09 depending on the specific problem.

### B. Performance of the GPU offloading only scheme

Second, we present our initial GPU-based experiments, in which our GPU-only (i.e. considering the portion variable $\theta$ equal to one: $\theta=1$) CUDA-based scheme is compared to the CPU-only implementation for varying number of cores and for both GPUs (either alone or in parallel / concurrent use). The performance gains achieved by our GPU-only approach are shown in Tables III, IV and V as well as in Fig. 1. Specifically, in Table III the corresponding results for the Rtx 2080Ti are given, in Table IV the results for the Titan Rtx, and in Table V the results for both the gpus working in parallel. Later on we present the additional performance gains achieved by our CPU/GPU collaboration scheme.

The experimental results given in Tables III, IV and V have been extracted over a randomly generated dense linear problem of size equal to 25000x25000 and structural properties similar to the ones of [19,33]. The size of the specific problem is the largest in the whole set of our experiments and it maximizes the speedup values for all the measured cases. It is also close to the maximum size that may be stored and processed comfortably within the memory of the Rtx 2080Ti card (which is equal to 11GB and corresponds to the weaker among the two cards involved in our experiments).

In the first two columns of each table, the obtained experimental values for our OpenMP-based (using the CPU only) implementation are given. More concretely, relevant experiments have been run for different number of cores (up to 32), and the execution time as well as the speedup ($Sp$) values achieved are presented per iteration. In addition, in the next two columns the execution time as well as the additional speedup achieved per iteration by our Cuda-based (using the GPUs only) implementation with the Rtx 2080Ti GPU (Table III), the Titan Rtx gpu (Table IV), and both the GPUs working in parallel (Table V) are presented. Note also that with respect to the use of both GPUs in parallel (Table V), the distribution coefficients $\theta_1$ and $\theta_2$ have been selected to be equal to $\theta_1=0.450$ and $\theta_2=0.550$, for Rtx 2080Ti and Tital Rtx respectively. The above distribution is almost equivalent to the processing capabilities of the two GPUs and leads to the best cooperative performance (as it was measured in related experiments with other possible alternatives). Determining the most effective coefficient values is even more important in our next set of experiments given in section V.C where all the underlying components (CPU and GPUs) work in parallel.

As it can be seen in Table III an additional speed-up of 2.34 is achieved with the use of the Rtx 2080Ti card, over the full configuration (32 cores) of the underlying multicore processing platform. Moreover, this additional performance gain becomes equal to 2.76 when the Titan Rtx card is used, whereas it becomes equal to 4.13 when both the cards are suitably used in parallel multi-GPU configuration. The above speedup values are surely satisfactory and they certify the added value offered by the use of such desktop-level GPUs for intensive scientific computing. Also, they are highly competitive to similar attempts in the literature, and certainly better comparing to the ones of [20]. Furthermore, the results of Table V clearly demonstrate an excellent performance, especially in the case the two cards work in parallel/concurrent mode, since their performance is compared to a very powerful multicore computer and over a extremely heavy test case.

Furthermore, in Fig.2 the performance of our multi-GPU scheme (i.e. using the Titan Rtx and the Rtx 2080Ti cards working in parallel over test problems of varying size), in terms of the speedup gained in each case is graphically presented. Dense LP problems generated at random (with sizes from 640x640 to 25000x25000) were used for the completion of our experiments; all with similar properties like in [19,33]. It must also be noted that the corresponding speedup measurements have been extracted comparing to the sequential implementation executing in one core. As it was normally expected, as the problem size increases the speedup values increase too. Also, the achieved speedup is observed to reach a quite high value (close to the maximum) for test problems equal to 2500x2500 or grater. On the other hand, it looks to decrease steadily for LP problems which are smaller in size. This is due to the fact that when the size of the problem decreases the processing load assigned to each computing component also decreases analogously. As a result, the CPU-GPU communication overhead as well as the processing overhead of the reduction procedures executed in GPUs, tend to be the dominant factors with respect to the total execution time.

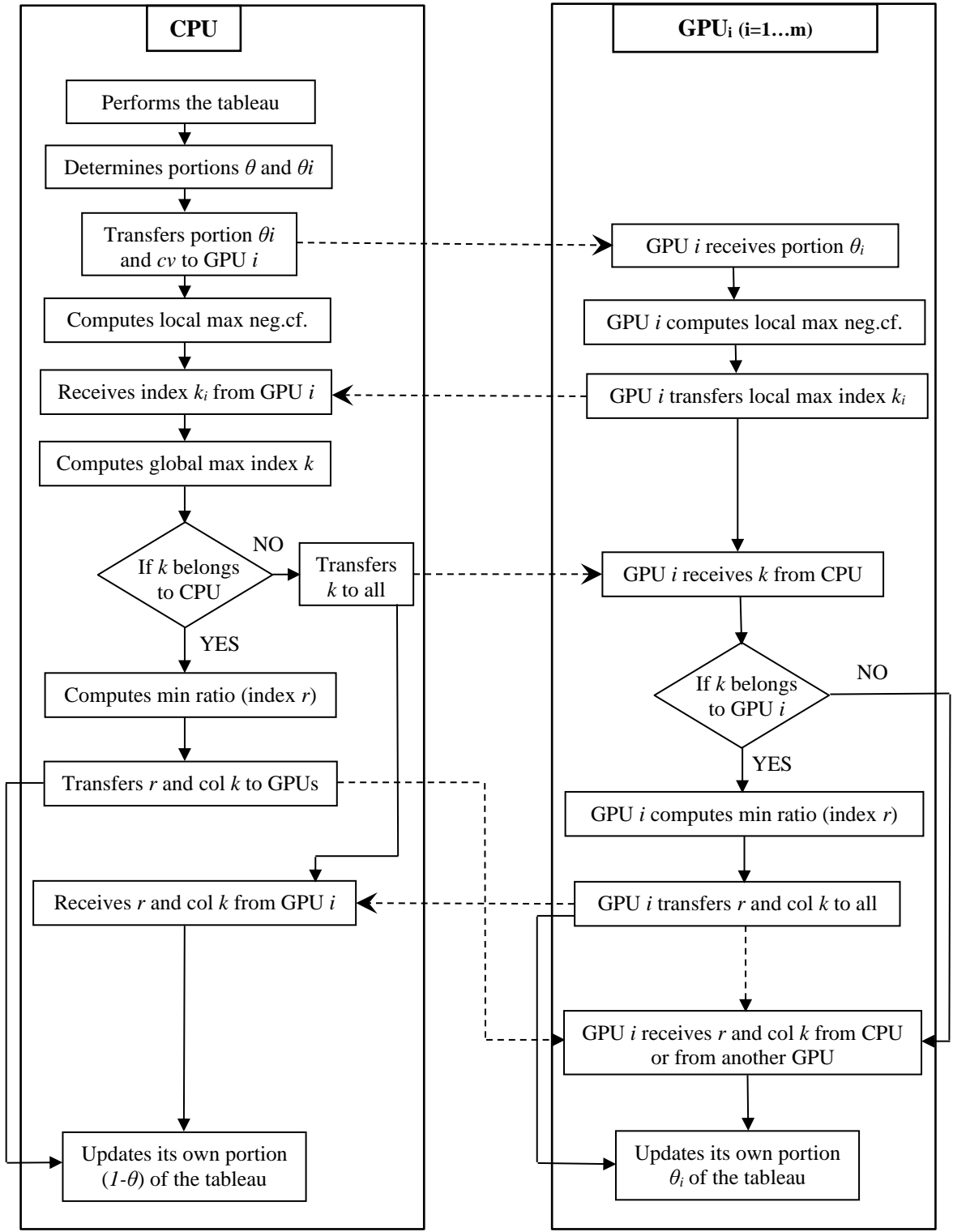

Fig. 1. Implementation flow of the algorithm for one iteration (the solid line arrows imply execution flow – the dashed line arrows imply data transfer)

TABLE II. EFFICIENCY OF THE BASIC OPENMP SCHEME FOR VERY LARGE PROBLEMS

| Linear Problems | Speed-up & Efficiency / OpenMP | | | | | | | |
|---|---|---|---|---|---|---|---|---|
| | 4 cores | | 8 cores | | 16 cores | | 32 cores | |
| | Sp | Ep | Sp | Ep | Sp | Ep | Sp | Ep |
| **FIT2P (3000x13525)** | 3.96 | 98.99% | 7.85 | 98.09% | 15.40 | 96.27% | 30.09 | 94.03% |
| **80BAU3B (2263x9799)** | 3.93 | 98.24% | 7.77 | 97.08% | 15.11 | 94.43% | 29.21 | 91.29% |
| **QAP15 (6330x22275)** | 3.91 | 97.74% | 7.67 | 95.82% | 14.65 | 91.59% | 27.91 | 87.23% |
| **MAROS-R7 (3136x9408)** | 3.89 | 97.23% | 7.59 | 94.82% | 14.34 | 89.64% | 27.06 | 84.56% |
| **QAP12 (3192x8856)** | 3.88 | 96.98% | 7.55 | 94.31% | 14.14 | 88.36% | 26.47 | 82.72% |
| **DFL001 (6071x12230)** | 3.87 | 96.73% | 7.55 | 94.31% | 14.22 | 88.87% | 26.73 | 83.53% |
| **GREENBEA (2392x5405)** | 3.86 | 96.48% | 7.46 | 93.24% | 13.85 | 86.57% | 25.84 | 80.76% |
| **STOCFOR3 (16675x15695)** | 3.81 | 95.22% | 7.32 | 91.46% | 13.37 | 83.59% | 24.75 | 77.34% |

### C. Performance of the hybrid CPU/GPU scheme

In the next part of our experimental evaluation, the performance of our hybrid CPU/GPU collaborative approach, and its superiority when compared to the native GPU-based approach is shown. The corresponding results were extracted by setting factor $\theta$ (which controls the distribution of the computational load) to the appropriate varying values; i.e ranging from 0 (for the CPU-based only scheme) to 1 (for the GPU-based only scheme) in steps of 0.1. Also, in the case of multi-GPU configuration the portion $\theta$ is further spread to the load distribution factors $\theta_1$ and $\theta_2$ for the Rtx 2080Ti and Titan Rtx cards respectively, according to their processing capabilities.

The relevant execution times are presented in tables VI, VII and VIII. First, the corresponding times supposing the each one of the involved GPUs works alone are given in tables VI and VII, whereas in Table VIII the results obtained supposing the two GPUs work in parallel are demonstrated. In the latter case, the two GPUs are assumed to share the computational load specified for GPU offloading adequately (according to the varying value of θ which spreads to θ1 and θ2 respectively). In all the experiments the more representative cases of 16 and 32 CPU collaborative cores are considered. The extracted results were obtained using the test LP problem of size 25000x25000 (generated at random) referred in the previous paragraph. The reader may notice that in all the different runs, for at least one value of $\theta$ the achieved execution time is better than in the GPU-based only approach. The corresponding time improvements should especially be noticed by staring at the last two columns of each table, for large values of $\theta$. Based on the above observations, the added value of using all the computing resources concurrently (i.e. the CPU in collaboration with both the GPUs – instead of using only the GPU resources, either alone or together in parallel), is clearly validated.

More concretely, the maximum improvement has been achieved – as it was expected – for 32 CPU-cores (the maximum multicore configuration used in our experiments), where we have an improvement of 22.5% (from 0.2044 to 0.1584) for the Rtx 2080Ti GPU, 19.4% (from 0.1732 to 0.1396) for the Titan Rtx GPU, and 11.6% (from 0.1157 to 0.1023) when both the GPUs are used in parallel. In terms of speedup values, the above mentioned improvements imply an additional speedup improvement of up to 1.29, 1.24 and 1.13 (comparing to the GPU-only scheme) when using Rtx 2080Ti, Titan Rtx, and both of them, respectively. These achievements are comparable to the ones presented in other recent works in the literature [11,14,16-17], and also they are the first ones achieved with multi-GPU configuration (which makes the CPU component less competitive).

Furthermore, the value of $\theta$ that leads to the larger improvement for 16 CPU-cores is 0.8 for the Rtx 2080Ti, as well as for the Titan Rtx, whereas when both of them are used in parallel it becomes equal to 0.9. The corresponding value of $\theta$ for 32 CPU-cores becomes 0.7 for the Rtx 2080Ti and Titan Rtx GPUs, and equal to 0.8 when both of them are used in parallel. As it was expected, all the above values are over 0.5, due to the fact that the GPU-based only approach could lead itself to a clear speedup (over the CPU-based onlyapproach) even if it's applied without the CPU collaboration. As a result, it may easily noticed that if we want to obtain the best possible performance for our hybrid collaboration scheme, we would definitely offload the larger fraction of the simplex tableau to the (one or more) GPU components of the computing platform. Moreover, the achieved improvement naturally increases as the number of the CPU cores increases, due to the fact that a larger fraction of the total work load is being performed by the CPU without performance degradation.

TABLE III.    SPEEDUP FOR 2080Ti GPU IMPLEMENTATION

| P | CPU (multi-threaded) | | GPU (Rtx 2080Ti) | |
|---|---|---|---|---|
| #cores | Time/iter | Sp | Time/iter | Sp (plus) |
| 4 | 3.1751 | 3.78 | 0.2044 | 15.53 |
| 8 | 1.6139 | 7.44 | 0.2044 | 7.90 |
| 12 | 1.2235 | 9.81 | 0.2044 | 5.99 |
| 16 | 0.8874 | 13.52 | 0.2044 | 4.34 |
| 20 | 0.7145 | 16.79 | 0.2044 | 3.50 |
| 24 | 0.6025 | 19.92 | 0.2044 | 2.95 |
| 28 | 0.5319 | 22.56 | 0.2044 | 2.60 |
| 32 | 0.4773 | 25.14 | 0.2044 | 2.34 |

TABLE IV.    SPEEDUP FOR TITAN RTX GPU IMPLEMENTATION

| P | CPU (multi-threaded) | | GPU (Titan Rtx) | |
|---|---|---|---|---|
| #cores | Time/iter | Sp | Time/iter | Sp (plus) |
| 4 | 3.1751 | 3.78 | 0.1732 | 18.33 |
| 8 | 1.6139 | 7.44 | 0.1732 | 9.32 |
| 12 | 1.2235 | 9.81 | 0.1732 | 7.06 |
| 16 | 0.8874 | 13.52 | 0.1732 | 5.12 |
| 20 | 0.7145 | 16.79 | 0.1732 | 4.13 |
| 24 | 0.6025 | 19.92 | 0.1732 | 3.48 |
| 28 | 0.5319 | 22.56 | 0.1732 | 3.07 |
| 32 | 0.4773 | 25.14 | 0.1732 | 2.76 |

TABLE V.    SPEEDUP FOR MULTI-GPU IMPLEMENTATION

| P | CPU (multi-threaded) | | 2xGPU (Titan + 2080Ti) | |
|---|---|---|---|---|
| #cores | Time/iter | Sp | Time/iter | Sp (plus) |
| 4 | 3.1751 | 3.78 | 0.1157 | 27.44 |
| 8 | 1.6139 | 7.44 | 0.1157 | 13.95 |
| 12 | 1.2235 | 9.81 | 0.1157 | 10.57 |
| 16 | 0.8874 | 13.52 | 0.1157 | 7.67 |
| 20 | 0.7145 | 16.79 | 0.1157 | 6.18 |
| 24 | 0.6025 | 19.92 | 0.1157 | 5.21 |
| 28 | 0.5319 | 22.56 | 0.1157 | 4.60 |
| 32 | 0.4773 | 25.14 | 0.1157 | 4.13 |

TABLE VI.    EXECUTION TIMES FOR CPU+2080Ti IMPLEMENTATION

| portion | CPU + GPU (Rtx 2080Ti) | | | |
|---|---|---|---|---|
| ($\theta$) | ($\theta 1$) | ($\theta 2$) | 16 cores | 32 cores |
| 0.0 | 1.000 | 0.000 | 0.8874 | 0.4773 |
| 0.1 | 0.900 | 0.000 | 0.8345 | 0.4633 |
| 0.2 | 0.800 | 0.000 | 0.7228 | 0.4125 |
| 0.3 | 0.700 | 0.000 | 0.6187 | 0.3791 |
| 0.4 | 0.600 | 0.000 | 0.5134 | 0.3068 |
| 0.5 | 0.500 | 0.000 | 0.3905 | 0.2452 |
| 0.6 | 0.400 | 0.000 | 0.2967 | 0.1961 |
| 0.7 | 0.300 | 0.000 | 0.2171 | 0.1584 |
| 0.8 | 0.200 | 0.000 | 0.1788 | 0.1713 |
| 0.9 | 0.100 | 0.000 | 0.1942 | 0.1897 |
| 1.0 | 0.000 | 0.000 | 0.2044 | 0.2044 |

TABLE VII.    EXECUTION TIMES FOR CPU+Titan Rtx IMPLEMENTATION

| portion | CPU + GPU (Titan Rtx) | | | |
|---|---|---|---|---|
| ($\theta$) | ($\theta 1$) | ($\theta 2$) | 16 cores | 32 cores |
| 0.0 | 0.000 | 1.000 | 0.8874 | 0.4773 |
| 0.1 | 0.000 | 0.900 | 0.8123 | 0.4526 |
| 0.2 | 0.000 | 0.800 | 0.6908 | 0.3936 |
| 0.3 | 0.000 | 0.700 | 0.5765 | 0.3203 |
| 0.4 | 0.000 | 0.600 | 0.4768 | 0.2726 |
| 0.5 | 0.000 | 0.500 | 0.3674 | 0.2115 |
| 0.6 | 0.000 | 0.400 | 0.2561 | 0.1695 |
| 0.7 | 0.000 | 0.300 | 0.1873 | 0.1396 |
| 0.8 | 0.000 | 0.200 | 0.1588 | 0.1535 |
| 0.9 | 0.000 | 0.100 | 0.1634 | 0.1602 |
| 1.0 | 0.000 | 0.000 | 0.1732 | 0.1732 |

TABLE VIII.    EXECUTION TIMES FOR CPU+2xGPU IMPLEMENTATION

| portion | CPU + 2xGPU (Titan + 2080Ti) | | | |
|---|---|---|---|---|
| ($\theta$) | ($\theta 1$) | ($\theta 2$) | 16 cores | 32 cores |
| 0.0 | 0.000 | 0.000 | 0.8874 | 0.4773 |
| 0.1 | 0.045 | 0.055 | 0.8332 | 0.4526 |
| 0.2 | 0.090 | 0.110 | 0.7552 | 0.3832 |
| 0.3 | 0.135 | 0.165 | 0.6004 | 0.3214 |
| 0.4 | 0.180 | 0.220 | 0.4793 | 0.2526 |
| 0.5 | 0.225 | 0.275 | 0.3624 | 0.1915 |
| 0.6 | 0.270 | 0.330 | 0.2643 | 0.1455 |
| 0.7 | 0.315 | 0.385 | 0.1891 | 0.1199 |
| 0.8 | 0.360 | 0.440 | 0.1168 | 0.1023 |
| 0.9 | 0.405 | 0.495 | 0.1102 | 0.1089 |
| 1.0 | 0.450 | 0.550 | 0.1157 | 0.1157 |

Additionally, the reader may notice in the second and the third column of Table VIII, the balance followed with regard to the load distribution factors $\theta_1$ and $\theta_2$ when both the Rtx 2080Ti and the Titan Rtx GPUs are used in parallel. As it can be seen, (a) in the case of the comparison to 16-core CPU, $\theta_1$ and $\theta_2$ take the values 0.405 and 0.495 respectively (thus forming a total portion $\theta$ equal to 0.9 and leaving a portion 1-$\theta$ equal to 0.1 for CPU processing), whereas (b) in the case of the comparison to 32-core CPU, $\theta_1$ and $\theta_2$ take the values 0.360 and 0.440 respectively (thus forming a total portion $\theta$ equal to 0.8 and leaving a portion 1-$\theta$ equal to 0.2 for CPU processing). In both cases the selection of the partial load distribution factors ($\theta_1$ and $\theta_2$), has been based on the processing capabilities of the two GPUs, and actually it is proved to lead to the best cooperative performance (as it comes out from the measurements taken for many different test values).

The overall performance of our hybrid approach (with either one or multiple GPUs) for LP problems of different sizes is presented in Fig.3. More concretely, the relevant speedup curves are shown for the test LP problems generated at random with sizes 640x640 to 25000x25000, along with the speedup curves of Fig.2.

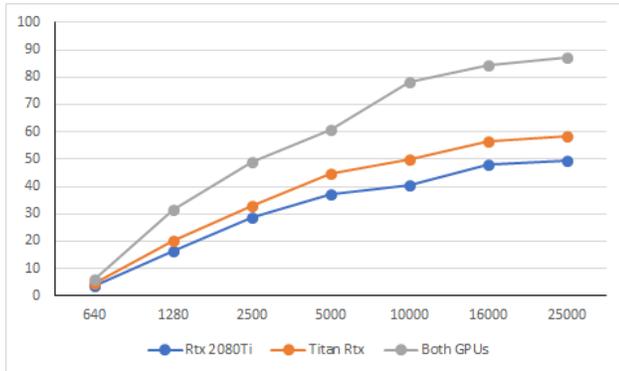

Fig. 2. Speed-up curves for different LP sizes

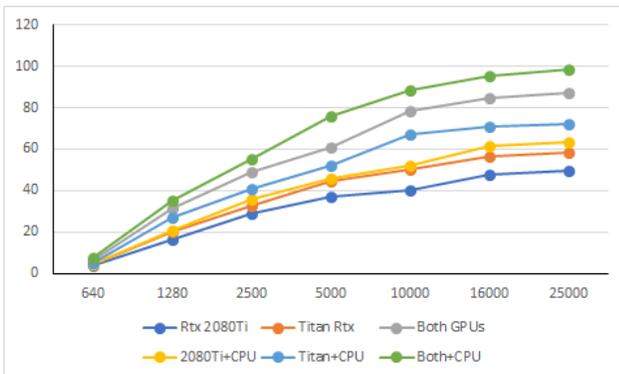

Fig. 3. Speedup curves for CPU/GPU collaboration

The corresponding values have been extracted for 32 CPU cores and by setting the value of $\theta$ in the appropriate level that maximizes the performance in every case. Beyond the validation of the highly effective behavior of the proposed hybrid scheme, it should also be noted that as the problem size decreases the speedup values decrease too, considering both the tested GPUs (either each one alone or working together). A natural explanation of this behavior lies in the fact that the distributed work load decreases analogously as the size of the problem decreases too, and also, in this case, the CPU to GPU communication overhead finally becomes much more crucial with respect to the total execution time.

## VI. Conclusion

A novel hybrid multi-GPU parallel approach with highly effective behavior is presented throughout the paper, with respect to the standard full tableau simplex algorithm. It has also been extensively evaluated in terms of typical performance measures, like the well-known *speedup* and *efficiency* computed values. The proposed CPU/GPU collaboration scheme, has been designed, implemented and evaluated using a hybrid high performance computing platform of the real world (a powerful 32-core multicore machine with two high-spec GPUs), and involving the well-known standard Netlib test LP problems (as the basic test case) in combination with a number of randomly generated LP problems of very large size. The corresponding implementations were based on the well-known and broadly accepted OpenMP and CUDA parallelization frameworks.

The extracted experimental results demonstrate the superiority of the proposed hybrid CPU/GPU collaboration scheme (when compared to the prior highly effective solution of a GPU-only implementation in almost all cases), which validates the added value of using both resources concurrently. The GPU-only implementation has also been proved to be comparable to other related works in the bibliography, as well as clearly superior to a pure CPU-only implementation. Specifically, the GPU-only implementation leads to speedup values of up to 2.34, 2.76 and 4.13 (comparing to the CPU-only implementation with 32 cores) when using the Rtx 2080Ti, the Titan Rtx and both of them, respectively. Moreover, the hybrid GPU/GPU implementation leads to an additional speedup of up to 1.29, 1.24 and 1.13 (comparing to the corresponding GPU-only implementation) when using the Rtx 2080Ti, the Titan Rtx and both of them, respectively. The above achievements are very satisfactory and comparable to the ones presented in other recent works in the literature [11,14,16-17]. Moreover, they are the first ones achieved with multi-GPU configuration, which makes the CPU component less competitive.

### References

[1] K. Murty. Linear Programming. Wiley, New York, 1983.

[2] J.A. Hall. Towards a Practical Parallelization of the Simplex Method. Computational Management Science, Springer, 7(2), 2010, pp. 139-170.

[3] E.S. Badr, M. Moussa, K. Paparrizos, N. Samaras and A. Sifaleras. Some Computational Results on MPI Parallel Implementation of Dense Simplex Method. World Acad. of Science, Engineering, Technology, 23,2008, pp. 778-781.

[4] J. Qin and D.T. Nguyen. A Parallel-vector Simplex Algorithm on Distributed-Memory Computers. Structural Optimizations, 11(3), 1996, pp. 260-262.

[5] G. Yarmish and R.V. Slyke. A Distributed Scaleable Simplex Method. Journal of Supercomputing, Springer, 49(3), 2009, pp. 373-381.

[6] M. Lubin, J.A. Hall, C.G. Petra and M. Anitescu. Parallel Distributed- Memory Simplex for Large-Scale Stochastic LP Problems. Computational Optimization and Applications, 55(3), 2013, pp. 571- 596.

[7] K. Sivaramakrishnan. A Parallel Interior Point Decomposition Algorithm for Block Angular Semidefinite Programs. Comput Optim Appl J, Springer, 46(1), 2010, pp. 1-29.

[8] T. Hoefler, J. Dinan, D. Buntinas, P. Balaji, B. Barrett, R. Brightwell, W.D. Gropp, V. Kale, R. Thakur. MPI + MPI: A New Hybrid Approach to Parallel Programming with MPI Plus Shared Memory. Computing J, Springer, 95, 2013, pp. 1121-1136.

[9] R. van der Pas, E. Stotzer and C. Terboven. Using OpenMP – The Next Step. MIT Press, 2017.

[10] R. Rabenseifner, G. Hager and G. Jost. Hybrid MPI and OpenMP Parallel Programming. Supercomputing 2013 Conference, Nov 17-22, Denver, USA, Tutorial, http://openmp.org/wp/sc13-tutorial-hybrid-mpi-and-openmp-parallel-programming, 2013.

[11] B. Mamalis and M. Perlitis, "Simplex Parallelization in a Fully Hybrid Hardware Platform", in International Journal of Advanced

Computer Science and Applications (IJACSA), Vol. 8, No. 4, pp. 356-365, 2017.

[12] B. Mamalis and G. Pantziou. Advances in the Parallelization of the Simplex Method. Proceedings of ALGO 2015 Annual Event, Springer, LNCS 9295 Festschrift (P. Spirakis), September 14-18, Patras, Greece, 2015, pp. 281-307.

[13] O. Adenikinju, J. Gilyard, J. Massey and T. Stitt. Concurrent Solutionsto Linear Systems using Hybrid CPU/GPU Nodes, SIAM UndergraduateResearch Online, vol. 8, 2015, pp. 1-10.

[14] G. Bosilca, A. Bouteiller, A. Danalis, T. Herault, P. Lemarinier and J. Dongarra. DAGuE: A generic distributed DAG engine for High Performance Computing, Parallel Computing, 38 (1-2), 2012, pp. 37-51.

[15] G.F. Diamos and S. Yalamanchili. Harmony: an execution model and runtime for heterogenous many core systems, in 17th International Symposium On High performance distributed computing, HPDC'08, ACM, 2008, pp. 197-200.

[16] M. Fatica. Accelerating linpack with CUDA on heterogenous clusters, in 2nd Workshop on General Purpose Processing on Graphics Processing Units, GPGPU-2, ACM, New York, NY, USA, 2009, pp. 46-51.

[17] J. Lang and G. Runger. Dynamic distribution of workload between CPU and GPU for a parallel conjugate gradient method in an adaptive FEM, ICCS 2013 Conference, in Procedia Computer Science, 18, 2013, pp. 299-308.

[18] B. Mamalis and M. Perlitis. Hybrid Parallelization of Standard Full Tableau Simplex Method with MPI and OpenMP. Proceedings of the 18th Panhellenic Conference in Informatics, ACM ICPS, 2014, pp. 1-6.

[19] M.E. Lalami, V. Boyer and D. El-Baz. Efficient Implementation of the Simplex Method on a CPU-GPU System, IEEE International Parallel & Distributed Processing Symposium, 2011, pp. 1994-2001.

[20] B. Mamalis and M. Perlitis. A Hybrid Parallelization Scheme for Standard Simplex Method based on CPU/GPU Collaboration, Proceedings of the 20th Panhellenic Conference in Informatics, ACM ICPS, 2016, pp. 12.

[21] J.A. Hall and K. McKinnon. ASYNPLEX an Asynchronous Parallel Revised Simplex Algorithm. Annals of Operations Research, 81, 1998, pp. 27-49.

[22] W. Shu and M.Y. Wu. Sparse Implementation of Revised Simplex Algorithms on Parallel Computers. Proceedings of the 6th SIAM Conference in Parallel Processing for Scientific Computing, Norfolk, VA, USA, 1993, pp. 501-509.

[23] M.E. Thomadakis and J.C. Liu. An Efficient Steepest-edge Simplex Algorithm for SIMD Computers. Proceedings of the Intl. Conference on Supercomputing, Philadelphia, PA, USA, 1996, pp. 286-293.

[24] J. Eckstein, I. Boduroglu, L. Polymenakos and D. Goldfarb. Data-Parallel Implementations of Dense Simplex Methods on the Connection Machine CM-2. ORSA Journal on Computing, Vol. 7 (4), 1995, pp. 402-416.

[25] C.B. Stunkel. Linear Optimization via Message-based Parallel Processing. Proceedings of Intl. Conf. on Parallel Processing, Pennsylvania, PA, USA, 1998, pp. 264-271.

[26] D. Klabjan, L.E. Johnson and L.G. Nemhauser. A Parallel Primal-dual Simplex algorithm.. Operations Research Letters, Vol 27 (2), 2000, pp. 47-55.

[27] I. Maros and G. Mitra. Investigating the Sparse Simplex Method on a Distributed Memory Multiprocessor, Parallel Computing, 26(1), 2000, pp. 151-170.

[28] J.A. Hall and Q. Huangfu. A high performance dual revised simplex solver. Parallel Processing and Applied Mathematics, LNCS 7203, Springer, 2012, pp. 143-151.

[29] Q. Huangfu and J.A. Hall. Parallelizing the dual revised simplex method. Technical Report ERGO-14-011, http://www.maths.ed.ac.uk/ hall/Publications.html, 2014.

[30] B. Mamalis, G. Pantziou, D. Kremmydas and G. Dimitropoulos. Highly Scalable Parallelization of Standard Simplex Method on a Myrinet Connected Cluster Platform. Acta Intl. Journal of Computers and Applications, 35(4), 2013, pp. 152-161.

[31] D.G. Spampinato and A.C. Elster. Linear Optimization on Modern GPUs. In Proc. of the 23rd IEEE IPDPS'09 International Conference, 2009, pp. 1-8.

[32] J. Bieling, P. Peschlow and P. Martini. An efficient GPU implementation of the revised Simplex method. In Proc. of IEEE 24th International Symposium on the Parallel & Distributed Processing, Workshops and Phd Forum (IPDPSW), 2010, pp. 1-8.

[33] M.E. Lalami, D. El-Baz and V. Boyer. Multi GPU implementation of the simplex algorithm, in Proc. of the 2011 IEEE 13th International Conference on High Performance Computing and Communications (HPCC), Banff, Canada, 2011, pp. 179-186.

[34] X. Meyer, P. Albuquerque and B. Chopard. A multi-GPU implementation and performance model for the standard simplex method, in Proc. of the 1st Intl. Symposium and 10th Balkan Conference on Operational Research, Thessaloniki, Greece, 2011, pp. 312–319.

[35] N. Ploskas and N. Samaras. Efficient GPU-based implementations of simplex type algorithms. Applied Mathematics and Computation, 250, 2015, pp. 552–570.

[36] V. Boyer and D. El-Baz. Recent Advances on GPU Computing in Operations Research. In Proc. of IEEE 27th International Symposium on Parallel & Distributed Processing Workshops and PhD Forum (IPDPSW), 2013, pp. 1778-1787.

[37] A. Gurung, B. Das and R. Rajarshi. Simultaneous Solving of Linear Programming Problems in GPU, in Proc. of IEEE HIPC 2015 Conference: Student Research Symposium on HPC, Vol. 8, Bengaluru, India, 2015, pp. 1-5.

[38] C.T. Yang, C.L. Huang and C.F. Lin. Hybrid CUDA, OpenMP, and MPI parallel programming on multicore GPU Clusters, Computer Physics Communications, 182, 2011, pp. 266–269.

[39] J. Lee, M. Samadi, Y. Park and S. Mahlke. Transparent CPU-GPU Collaboration for Data-Parallel Kernels on Heterogeneous Systems, in Proc. of the 22nd International Conference on Parallel Architectures and Compilation Techniques, PACT '13, 2013, pp. 245-256.

[40] Z. Zhong, M. Feng and D. Liu. Parallelization of Revised Simplex Algorithm on GPUs. In International Conference on Network and Information Systems for Computers, 2015, pp. 349-353.

[41] L. He, H. Bai, Y. Jiang, D. Ouyang, S. Jiang. Revised simplex Algorithm for Linear Programming on GPUs with CUDA. Multimedia Tools and Applications, 2018, 77, 30035-30050.

[42] K. Perumalla and M. Alam. Design Considerations for GPU-based Mixed Integer Programming on Parallel Computing Platforms. ICPP Workshops, 2021, 27:1-27:7.

[43] U. A. Shah, S. Yousaf, I. Ahmad, S. U. Rehman and M. O. Ahmad. Accelerating Revised Simplex Method Using GPU-Based Basis Update. In IEEE Access, 2020, Vol. 8, pp. 52121-52138.

[44] A. Gurung and R. Ray. Simultaneous Solving of Batched Linear Programs on a GPU. In Proc. of ACM/SPEC Int. Conf. Perform. Eng. (ICPE), 2019, pp. 59-66.

[45] A. R. Gahrouei and M. Ghatee. Effective Implementation of GPU-based Revised Simplex algorithm applying new memory management and cycle avoidance strategies. arXiv.org > math > arXiv:1803.04378, Amirkabir University of Technology, Tehran, Iran, 2019.